\documentclass{article}
\usepackage[margin=1in]{geometry}
\usepackage[utf8]{inputenc}
\usepackage{graphicx,wrapfig}
\usepackage[singlelinecheck=false]{caption}
\usepackage[colorlinks=true,linkcolor=black,citecolor=black,urlcolor=black]{hyperref}
\usepackage{authblk}

\usepackage[authoryear,sort&compress]{natbib}
\setcitestyle{open={(},close={)}}

\usepackage[symbol]{footmisc}

\usepackage{layouts}

\title{Effects of anterior temporal lobe resection on cortical morphology}

\author[1]{Karoline~Leiberg\footnote{Corresponding author. School of Computing, Newcastle University, 1 Science Square, NE4 5TG, Newcastle upon Tyne, United Kingdom \\ Email address: k.leiberg2@newcastle.ac.uk}}
\author[2]{Jane~de~Tisi}
\author[2,3]{John~S~Duncan}
\author[1,4]{Bethany~Little}
\author[1,4,5]{Peter~N~Taylor}
\author[5,6,7]{Sjoerd~B~Vos}
\author[2,8,9]{Gavin~P~Winston}
\author[10]{Bruno~Mota}
\author[1,4,5]{Yujiang~Wang\footnote{Email address: yujiang.wang@newcastle.ac.uk}}

\affil[1]{CNNP Lab (www.cnnp-lab.com), Interdisciplinary Computing and Complex BioSystems Group, School of Computing, Newcastle University, Newcastle upon Tyne, UK}
\affil[2]{Department of Clinical \& Experimental Epilepsy, UCL Queen Square Institute of Neurology, London, UK}
\affil[3]{Chalfont Centre for Epilepsy, Chalfont St Peter, UK}
\affil[4]{Faculty of Medical Sciences, Newcastle University, Newcastle upon Tyne, United Kingdom}
\affil[5]{Queen Square Institute of Neurology, University College London, Queen Square, London, UK}
\affil[6]{Neuroradiological Academic Unit, Department of Brain Repair and Rehabilitation, UCL, UK}
\affil[7]{Centre for Medical Image Computing, University College London, London, UK}
\affil[8]{MRI Unit, Epilepsy Society, Buckinghamshire, UK}
\affil[9]{Division of Neurology, Department of Medicine, Queen's University, Kingston, Ontario, Canada}
\affil[10]{metaBIO Lab, Instituto de Física, Universidade Federal do Rio de Janeiro (UFRJ), Rio de Janeiro, Brazil}

\providecommand{\keywords}[1]
{
  \small	
  \textbf{\textit{Keywords:}} #1
}

\date{\vspace{-5ex}}

\begin{document}

\maketitle

\begin{abstract}

Anterior temporal lobe resection (ATLR) is a surgical procedure to treat drug-resistant temporal lobe epilepsy (TLE). Resection may involve large amounts of cortical tissue. Here, we examine the effects of this surgery on cortical morphology measured in independent variables both near the resection and remotely.

We studied 101 individuals with TLE (55 left, 46 right onset) who underwent ATLR. For each individual we considered one pre-surgical MRI and one follow-up MRI 2 to 13 months after surgery. We used our newly developed surface-based method to locally compute traditional morphological variables (average cortical thickness, exposed surface area, and total surface area), and the independent measures $K$, $I$, and $S$, where $K$ measures white matter tension, $I$ captures isometric scaling, and $S$ contains the remaining information about cortical shape. Data from 924 healthy controls was included to account for healthy ageing effects occurring during scans. A SurfStat random field theory clustering approach assessed changes across the cortex caused by ATLR.

Compared to preoperative data, surgery had marked effects on all morphological measures. Ipsilateral effects were located in the orbitofrontal and inferior frontal gyri, the pre- and postcentral gyri and supramarginal gyrus, and the lateral occipital gyrus and lingual cortex. Contralateral effects were in the lateral occipital gyrus, and inferior frontal gyrus and frontal pole.

The restructuring following ATLR is reflected in widespread morphological changes, mainly in regions near the resection, but also remotely in regions that are structurally connected to the anterior temporal lobe. The causes could include mechanical effects, Wallerian degeneration, or compensatory plasticity. The study of independent measures revealed additional effects compared to traditional measures.

\end{abstract}

\keywords{Temporal lobe epilepsy, Cortical morphology, Structural MRI, Epilepsy surgery}

\section{Introduction}

Anterior temporal lobe resection (ATLR) is a common surgical procedure to treat drug-resistant temporal lobe epilepsy (TLE). It removes the anterior (3-3.5 cm) of the middle and inferior temporal gyri and uncus, amygdala, anterior hippocampus, and parahippocampal gyrus \citep{Foldvary-Schaefer2007}. 

The effect of surgery on the cortical morphology of the remaining brain could inform how neighbouring, or connected regions affect each other's morphology, and the processes of postoperative plasticity. Previous studies found widespread, bilateral changes to cortical thickness \citep{Elias2021,Li2022,Zhao2021,Galovic2020} or volume \citep{Pajkert2020}. However, these studies had relatively small sample sizes (between 12 and 56 subjects), and there is little overlap in regions identified as affected by morphological changes between the studies. In fact, some of these studies found conflicting effects, such as increased \citep{Elias2021} or decreased \citep{Galovic2020} cortical thickness in the contralateral anterior and middle cingulate cortex. ATLR has also been associated with altered white matter tract properties \citep{Concha2007,DaSilva2020,Faber2013,McDonald2010,Pustina2014,Winston2013}, particularly the ipsilateral uncinate fasciculus, inferior longitudinal, inferior fronto-occipital fasciculi, optic radiation, cingulum, fornix, and the corpus callosum; these alterations may also differ in individuals rendered seizure free and those experiencing seizures after surgery \citep{DaSilva2020}. Other studies have found widespread reductions in functional connectivity after surgery compared to healthy controls in connections that were not different to controls before surgery \citep{Morgan2020}, as well as dissimilar functional reorganisation for individuals rendered seizure free and individuals with recurrent seizures \citep{Liao2016}. Finally, 25-50\% of individuals with TLE experience significant declines in memory and language functions after ATLR \citep{Davies1998,Martin1998}, and postoperative changes in white matter are reported to be associated with recovery of language function \citep{Winston2013,Yogarajah2010}. However, in terms of brain morphology, the spatial characteristics of post operative changes and their relationship with seizure and cognitive outcomes are currently not clear. 

Traditional morphological analyses generally rely on measures such as cortical thickness, volume, and surface areas. However, these measures can yield conflicting results \citep{Alhusaini2012} and a single measure, such as volume, could be driven by multiple independent biological processes \citep{Panizzon2009}. Further, the measures of average cortical thickness ($T$), total surface area ($A_t$), and exposed surface area ($A_e$) covary tightly across species, individuals, and regions  \citep{Mota2015, Wang2016, Wang2019} according to a scaling law of cortical folding. More recently developed morphological measures were presented by \cite{Wang2021}. That study showed that traditional morphological measures can miss substantial information about cortical shape, and suggested alternative morphological measures of tension ($K$), isometric size ($I$), and shape ($S$). These morphological measures are theoretically and statistically independent, capture all morphological information in $T$, $A_t$ and $A_e$, and can detect morphological changes that might otherwise not be detected due to the covariance of the traditional variables.

In this study we examined 101 individuals with TLE who underwent ATLR. We performed a localised, surface-based analysis of cortical morphology to derive maps of cortical changes after ATLR using both traditional and recently developed morphology measures. Quantifying these changes after ATLR on a regional cortical basis could lead to a better understanding of the consequences and processes of reorganisation following ATLR, and the inter-relationships between connected regions.

\section{Methods}

\subsection{Independent morphological measures}\label{KIS}

\begin{figure}[h!]
\centering
\includegraphics[scale=0.6]{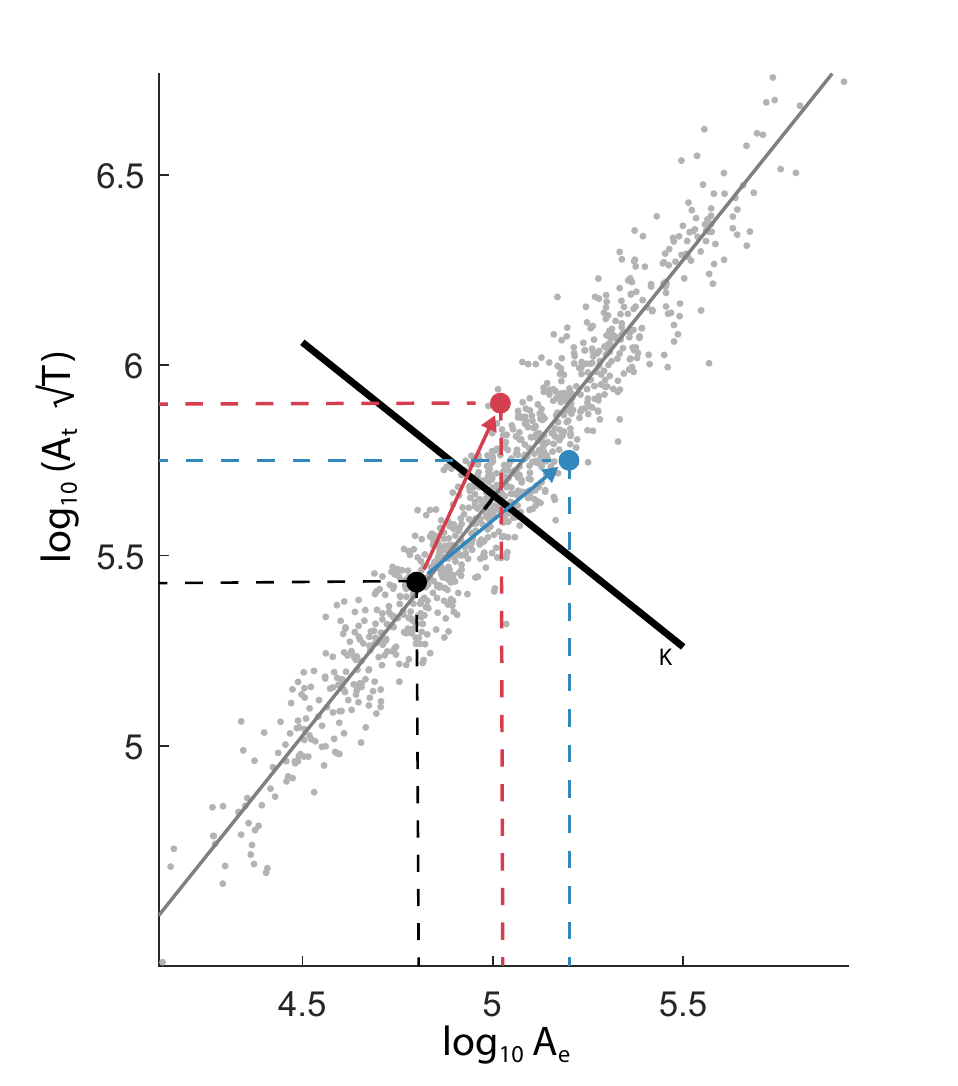}
\caption{\textbf{Schematic illustration with example data on how traditional measures can hide information due to their covariance, and how a change of coordinate system can reveal this information.} The plot shows a 2-dimensional projection of data points measured in thickness $T$, total surface area $A_t$, and exposed surface area $A_e$. Their agreement with the scaling law can be seen as the alignment with a slope of 1.25 (thin grey line). The two processes of changing from the initial black point to either blue or red appear similar on the original axes, both being increases in the same directions. It becomes apparent that they are in fact opposing processes when looking at changes along the new axis $K$ (thick black line).} \label{Fig6}
\end{figure}

We performed a morphological analysis in the recently developed framework \citep{Wang2021} to study cortical morphology based on the universal scaling law of cortical folding. This law states that across mammalian species \citep{Mota2015}, individual human brains \citep{Wang2016}, their lobes \citep{Wang2019}, and small local areas \citep{Leiberg2021}, the average cortical thickness $T$, the total surface area $A_t$, and the exposed surface area $A_e$ covary according to the equation
\begin{equation}\label{SL}
    A_t \sqrt T = k A_e^{1.25}.
\end{equation}
Here, $k$ is a constant that is relatively preserved across species compared to the range of variation in the traditional morphological measures $T$, $A_t$, and $A_e$, but shows consistent changes across age groups and between cohorts of healthy and diseased subjects \citep{Mota2015,Wang2016,Wang2019}.
This interdependence of the traditional morphological measures means that their interaction must be considered when studying effects in this framework. We use independent morphological measures that are derived from said scaling law as combinations of the traditional measures, and capture all morphological information they contain. Figure \ref{Fig6} shows a schematic example of how the interdependence of the traditional measures can hide information, which is revealed in the independent measures.

Importantly, the scaling law also directly challenges the direct use of these traditional morphological measures of surface and thickness, without consideration for their covariance. Instead an independent set of morphological measures is proposed, derived from the scaling law \citep{Wang2021}. Solving equation \ref{SL} for $\log k$ gives the first morphometric measure $K$ in this framework of describing cortical morphology. It is a dimensionless measure that we interpreted as the tension acting on the cortex \citep{Wang2016,Wang2019,Wang2021,Leiberg2021}. The second measure is $I$, a term of isometric size in all three traditional variables $T^2$, $A_t$, and $A_e$. $I$ is orthogonal to $K$. The final measure $S$ is the cross product of $K$ and $I$, so that all three are orthogonal to each other. $S$ contains all remaining information about cortical shape that is not captured in $K$ or $I$, and is a measure of complexity in cortical folding. The three new measures span orthogonal vectors in the three-dimensional morphological space of $T^2$, $A_t$, and $A_e$, and hence using $K$, $I$, and $S$ is simply a change of coordinate systems:

\begin{equation}\label{K}
    K = \log A_t + \frac{1}{4} \log T^2 - \frac{5}{4} \log A_e,
\end{equation}
\begin{equation}\label{I}
    I = \log A_t + \log T^2 + \log A_e,
\end{equation}
\begin{equation}\label{S}
    S = \frac{3}{2} \log A_t - \frac{9}{4} \log T^2 + \frac{3}{4} \log A_e.
\end{equation}

% Data
\subsection{Data \& preprocessing}

We used 3T T1-weighted structural magnetic resonance images (MRIs) of 101 individuals with TLE who underwent ATLR at the National Hospital for Neurology and Neurosurgery (NHNN). The analysis was carried out under approval by the Newcastle University Ethics Committee (8841/2020). For each subject we analysed an image acquired up to three years before surgery and one follow-up scan 2 to 13 months (median=3.6 months) after surgery. If patients had more than one follow-up scan, we used the one closest to surgery. MRI scans were taken under two different scanning protocols: 52 subjects were scanned with a resolution of 0.94x0.94x1.1mm (scanning protocol 1; for detailed acquisition parameters see  \cite{Nowell2016}) and the remaining 49 with 1x1x1mm (scanning protocol 2; for detailed acquisition parameters see  \cite{Vos2020}). Surgery outcomes were recorded 12 months after surgery according to the ILAE classification of seizure outcome \citep{Wieser2001}. We also used MR images of a cross-sectional cohort of 924 healthy controls, 22 of which were taken with scanning protocol 1 and 69 with scanning protocol 2 at the NHNN. The remaining 833 controls are subjects from the Nathan Kline Institute (NKI) data set \citep{Nooner2012}. Table \ref{table:1} shows the demographics of the data used.

\begin{table}[h!]
\centering
\begin{tabular}{p{6cm} p{3cm} p{3cm} p{3cm}} 
 \hline \\ [0.05ex]
  & Left TLE & Right TLE & Controls \\ [2ex] 
 \hline \\ [0.05ex] 
 Subjects (n) & 55 & 46 & 924 \\ [1.5ex]
 Sex (M/F) & 24/31 & 16/30 & 362/562 \\ [1.5ex]
 Median age at first scan in years (IQR) & 35.07 (17.82) & 35.58 (18.16) & 38 (36.5) \\ [1.5ex]
 Median age at surgery in years (IQR) & 36.45 (17.77)1 & 36.85 (19.40) & - \\ [1.5ex]
 Median postoperative interval in months (IQR) & 3.6 (1.52) & 3.6 (1.05) & - \\ [3ex]
 Median TLE duration in years (IQR) & 19 (27.99) & 21.5 (21.23) & - \\ [1.5ex]
 Seizure outcome group* 3+ (recurrent seizures) after one year (n) & 12 & 11 & - \\ [2ex]
 \hline
\end{tabular}
\caption{Demographics of individuals with TLE and controls. *ILAE scale for classification of outcome of epilepsy surgery}
\label{table:1}
\end{table}

MRI scans were preprocessed with the FreeSurfer \citep{Fischl2012} 6.0 recon-all pipeline, which produces a mesh representation of the grey matter surface, along with the grey matter thickness for each point on the grey matter surface. We manually corrected the grey- and white matter segmentation by using control points and corrected grey matter boundaries where necessary. The local gyrification index processing stream (\url{https://surfer.nmr.mgh.harvard.edu/fswiki/LGI}) was then used to acquire the smooth pial surface. This surface, which is obtained by closing sulci of the pial surface with a 15mm diameter sphere, is an outer surface wrapped tightly around the pial surface~ \citep{Schaer2008}.

We visually ensured that there were no distortions to the Desikan-Killiany atlas surface ROI labels around the resected area by comparing pre- and postoperative scans.

% Local method
\subsection{Computation of localised morphological measures}\label{Compute}

To analyse the effects of epilepsy surgery, we used a surface-based approach rather than a parcellation-based approach for increased sensitivity. We obtained surface-based morphological measures of the local cortical thickness, local surface area and local exposed area.

\begin{figure}[h!]
\centering
\includegraphics[width=16.5cm]{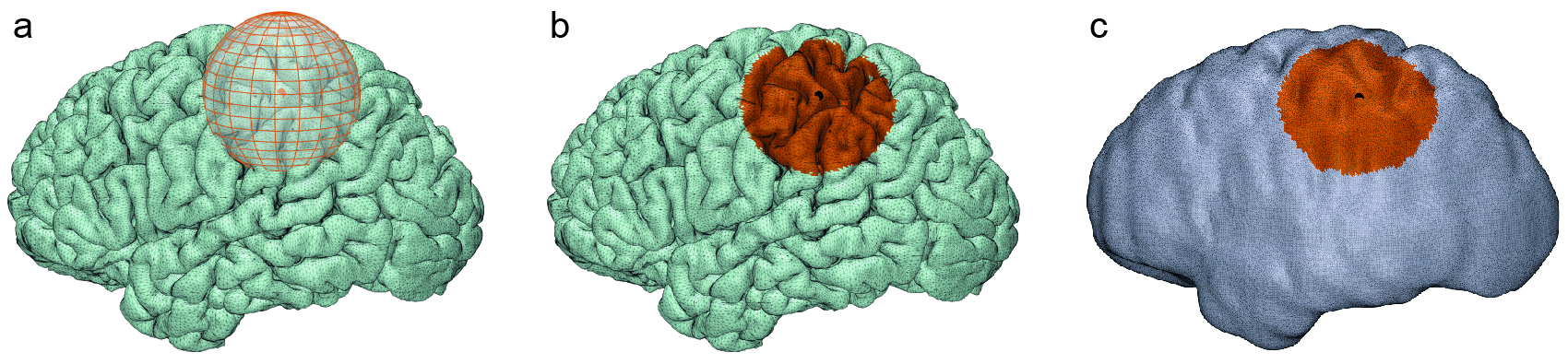}
\caption{\textbf{Method to extract local morphological measures.} a) Definition of neighbourhood of connected vertices within 3cm around each point. b) Computation of average thickness $^{p}T$ and total area $^{p}A_t$ of patch. c) Computation of exposed area $^{p}A_e$ of the patch.} \label{Fig1}
\end{figure}

We employed our previously published pipeline (fig. \ref{Fig1})  \citep{Leiberg2021}. In summary: we first downsampled the FreeSurfer pial surface to 5\% of its original density. We then defined a contiguous surface patch of 3~cm radius surrounding each point in the pial. We closed potential holes in the patch which can arise at the top of gyri or the bottom of sulci. We computed the pointwise average cortical thickness ($^{p}T$) and total surface area ($^{p}A_t$) from this patch around each point.

Next, we found a corresponding surface patch on the smooth pial surface by selecting all vertices on it for which their nearest vertex on the pial surface is part of the pial surface patch. From this, we calculated the exposed surface area ($^{p}A_e$) of the patch. We then fitted a convex hull over the patch on the smooth pial surface and calculated its integrated Gaussian curvature ($^{p} I_{G}$) as the sum of Gaussian curvatures of all points on the convex hull that are not lying on its edge. Since the integrated Gaussian curvature is conserved in closed surfaces, we used the proportion of curvature of a patch over the full pial surface to correct both surface areas $^{p}A_t$ and $^{p}A_e$ according to the reasoning laid out in previous publications \citep{Wang2019,Leiberg2021}: $^{p}A_t'= ^{p}A_t \times \frac{4 \pi}{^{p}I_{G}}$ and $^{p}A_e'=^{p}A_e \times \frac{4 \pi}{^{p}I_{G}}$.

The method of correcting surface areas does not perform well for regions that are particularly flat or located in deep sulci, so some vertices were missing data. Where possible, we imputed data on vertices that were missing values as the mean over their direct neighbouring vertices, but had to exclude the insula and the regions around the corpus callosum from our analysis, since these regions had entirely missing values. We then converted back to the full pial surface by using the value of the vertex on the downsampled pial surface that was closest to the pial surface.

To avoid the morphology of the resection affecting measures of nearby vertices, we excluded the entire ipsilateral temporal lobe from the pial surface before applying our method. We also applied this exclusion to the preoperative surfaces and processed all controls twice, once with the temporal lobes and once without, to keep the analysis consistent and comparable.

\subsection{Scaling law fit in data of individuals with TLE}

We assessed the fit of the scaling law to the cortical folding of individuals with TLE before surgery by regressing between the quantities $\log_{10} (^{p}A_e')$ and $\log_{10} (^{p}A_t' \sqrt{^{p}T})$ across each hemisphere, and comparing the slope of this regression to the predicted slope of $1.25$.

% GAMM
\subsection{Age, sex and scanning protocol correction}

After registering all subjects to the FreeSurfer fsaverage surface and performing a log transform, we performed a pointwise correction of the three traditional morphological measures $^{p}T$, $^{p}A_t'$, and $^{p}A_e'$ using generalised additive mixed models (GAMM), to account for differences due to age, sex, and scanning protocol. We did this with the R package mgcv (\url{https://CRAN.R-project.org/package=mgcv}). We trained a model with the formula

\begin{equation}
    y = s(\textrm{age}) \: \textrm{sex} + \beta \: \textrm{scanning protocol} + \epsilon,
\end{equation}

where $\beta$ is a model coefficient, $s$ is a smooth spline, $\epsilon$ is the error term, and $y$ is one of the measures $^{p}T$, $^{p}A_t'$, or $^{p}A_e'$, using restricted maximum likelihood as the method for smoothing parameter estimation. We trained this on all controls, with the large NKI data set ensuring we had robust models for the entire age range, and the site-specific controls matching the patients informing effects for the specific scanning protocols. We used the control data processed without/with the temporal lobes for correcting the ipsilateral/contralateral sides respectively by predicting the values of individuals with TLE both in pre- and postoperative data and using those predictions for correcting the individuals' data. Note how this automatically accounts for changes to the brain's morphology that occur due to healthy ageing between the time points of the pre- and postoperative scans.

% KIS
\subsection{Converting to independent morphological measures}

The traditional measures $^{p}T$, $^{p}A_t'$, and $^{p}A_e'$ were then converted to a set of independent morphological measures (section \ref{KIS}) in each patch, to account for the covariance in the traditional measures. We transformed the data, which had been corrected for the covariates age, sex, and scanning protocol, to the independent morphological measures $K, I,$ and $S$ according to their formulae
\begin{equation}\label{K}
    ^{p}K = \log (^{p}A_t') + \frac{1}{4} \log (^{p}T^2) - \frac{5}{4} \log (^{p}A_e'),
\end{equation}
\begin{equation}\label{I}
    ^{p}I = \log (^{p}A_t') + \log (^{p}T^2) + \log (^{p}A_e'),
\end{equation}
\begin{equation}\label{S}
    ^{p}S = \frac{3}{2} \log (^{p}A_t') - \frac{9}{4} \log (^{p}T^2) + \frac{3}{4} \log (^{p}A_e').
\end{equation}

We thus obtained six maps of cortical morphology for each subject's pre- and postoperative brain, measured in independent morphological measures ($^{p}K, ^{p}I, ^{p}S$) and traditional variables ($^{p}T$, $^{p}A_t'$, and $^{p}A_e'$) in a region for each point on the cortex.

For simplicity of notations, and as we will focus our analysis on patch-based measures, we will drop the indicator $^p$ for the patch, and in the following $K,I,S$ are the independent morphological measures of patches, and $T, A_t, A_e$ are the traditional measures of patches.

% Surfstat
\subsection{Surface effects}

We proceeded with an ipsilateral and contralateral analysis by combining individuals with left and right onset (see supplementary figures 1-4 for separate results for each onset site).

To allow for a pairwise analysis of the changes between the pre- and postoperative scans, we centred the data subject-wise at each vertex.

We used the Matlab toolbox SurfStat \citep{Worsley2009} for a surface based statistical analysis that corrects for multiple comparison and accounts for spatial correlation with random field theory. We employed a design matrix with effects for the group of preoperative surfaces and the group of postoperative surfaces and used a contrast to check for effects between the two categories. We repeated this for all six variables (traditional and independent), ending up with a surface map of the pairwise effect between the two scans for each. We applied a threshold for effects having a cluster-wise model significance of $p \leq 0.05$. We reported effects measured in Cohen's $f^2$.

% Other effects
\subsection{Covariate effects}

We tested for effects of the covariates: age at surgery, duration since first seizure, resection volume, side of resection, time between surgery and postoperative scan, and seizure outcome (ILAE 1\&2 vs 3+). We performed this by taking subject-wise differences between pre- and postoperative at each point and using a design matrix for each covariate separately to test for significance of the covariate on the effects of surgery.

\section{Results}

\subsection{Scaling law of cortical folding in individuals with TLE}

\begin{figure}[htp]
\centering
\includegraphics[scale=0.6]{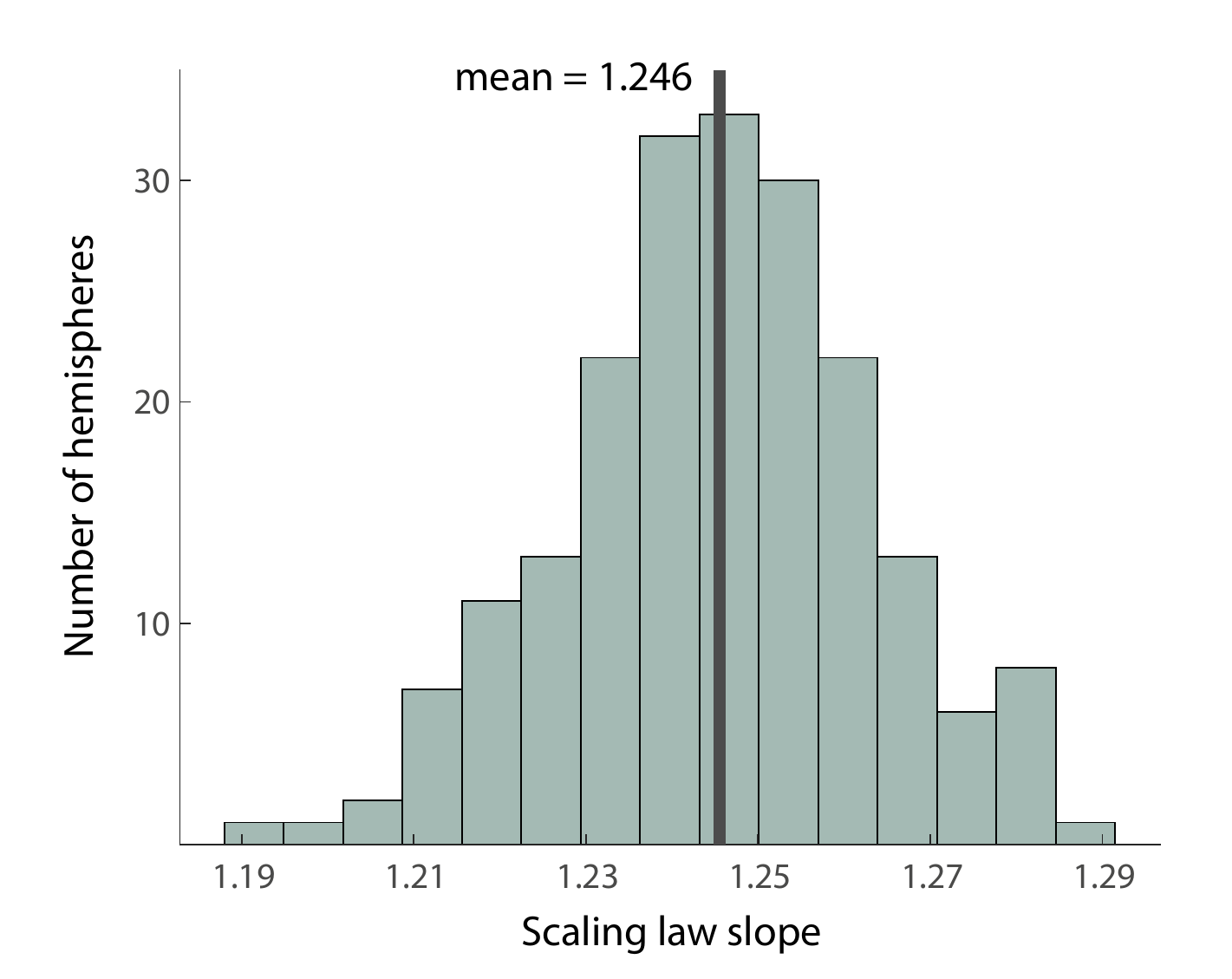}
\caption{\textbf{Distribution of observed scaling law slopes in individuals with TLE preoperatively.} Each subject's hemisphere is a single measurement of a slope and the distribution is formed across subjects. The mean of slopes is marked by the black line.} \label{Fig5}
\end{figure}

We first verified that the local folding in individuals with TLE follows the universal scaling law. As shown in Fig. \ref{Fig5}, the slopes of each subject's hemispheres are distributed around a mean of $1.246$, indicating that the scaling law also applies to our cohort of individuals with TLE.

\subsection{Morphological effects}

We found morphological effects in 5 main areas: 
\begin{itemize}
\item ipsilateral orbitofrontal and inferior frontal gyri
\item ipsilateral pre- and postcentral gyri and supramarginal gyrus
\item ipsilateral lateral occipital gyrus, and lingual cortex
\item contralateral lateral occipital gyrus
\item contralateral inferior frontal gyrus and frontal pole. 
\end{itemize}

\noindent In the following, we will go through changes in these areas by hemisphere and morphometric measure.

\subsubsection{Traditional morphological measures}

\begin{figure}[htp]
\centering
\includegraphics[width=16.5cm]{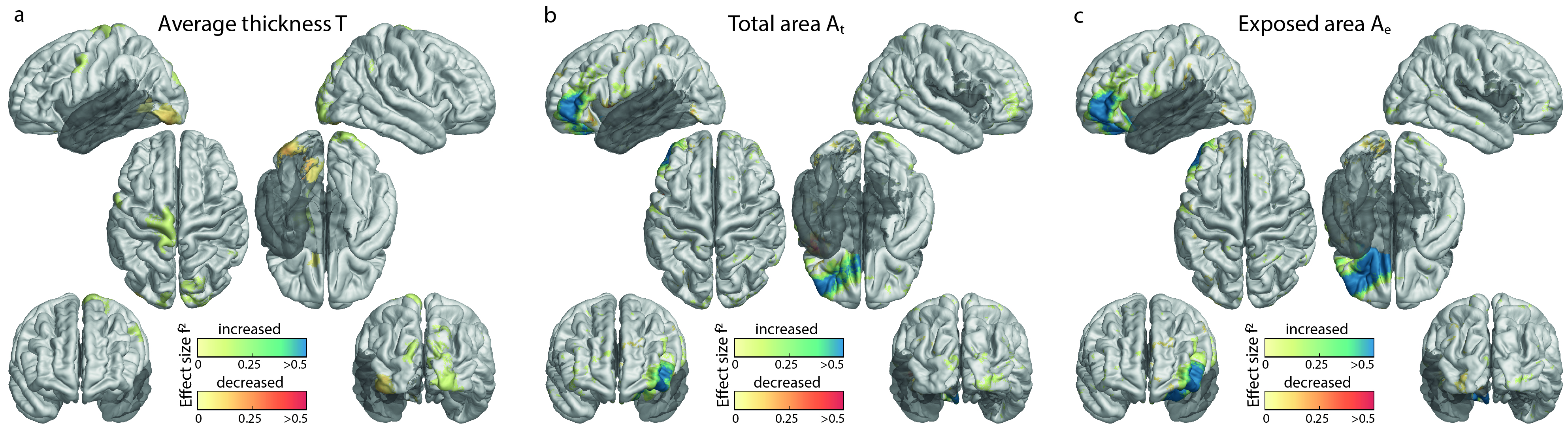}
\caption{Changes following ATLR in the morphological variables average cortical thickness $T$ (a), total surface Area $A_t$ (b), and exposed surface area $A_e$ (c). Effect clusters with statistical significance of $p \leq 0.05$ are shown and effect sizes are given in Cohen’s $f^2$ metric. Increases are in blue, decreases are in red. Regions excluded from the analysis, such as the temporal lobe that was operated upon, are in dark grey.} \label{Fig2}
\end{figure}

We first assessed effects of ATLR on the traditional measures of average cortical thickness, total surface area and exposed surface area. On the side ipsilateral to seizure onset, cortical thickness increased in the precentral gyrus and decreased in the lingual cortex and lateral occipital gyrus (fig. \ref{Fig2}a). The total surface area increased in the orbitofrontal and inferior frontal gyri (fig. \ref{Fig2}b). The exposed area also increased in the orbitofrontal and inferior frontal gyri, and reduced in the lingual cortex (fig. \ref{Fig2}c).

In the contralateral hemisphere, $T$ increased in the lateral occipital gyrus (fig. \ref{Fig2}a), whilst $A_t$ increased in the lateral occipital cortex and in the inferior frontal gyrus (fig. \ref{Fig2}b).

Average effect sizes for relevant regions can be found in supplementary section 1. See supplementary section 2.1 for effects by onset site.

\subsubsection{Independent morphological measures}

\begin{figure}[htp]
\centering
\includegraphics[width=16.5cm]{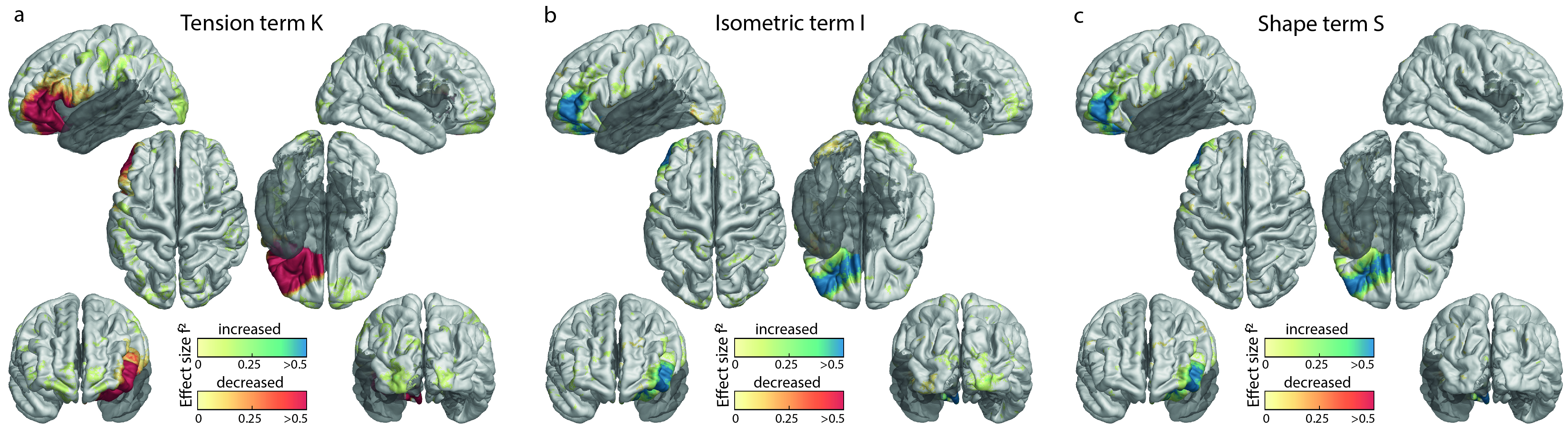}
\caption{Changes following ATLR in the independent morphological measures $K$ (a), $I$ (b), and $S$ (c). Effect clusters with statistical significance of $p \leq 0.05$ are shown and effect sizes are given in Cohen’s $f^2$ metric. Increases are in blue, decreases are in red. Regions excluded from the analysis, such as the temporal lobe that was operated upon, are in dark grey.} \label{Fig3}
\end{figure}

In the independent measures, on the ipsilateral side the pressure term $K$ decreased in the orbitofrontal and inferior frontal gyri, and increased in the pre- and postcentral and supramarginal gyri, as well as the lateral occipital cortex (fig. \ref{Fig3}a). The isometric term $I$ increased in the orbitofrontal and inferior frontal gyri, and decreased in the lateral occipital cortex (fig. \ref{Fig3}b). The shape term $S$ increased in the orbitofrontal and inferior frontal gyri (fig. \ref{Fig3}c).

In the contralateral hemisphere, $K$ increased in the lateral occipital cortex and in the frontal pole (fig. \ref{Fig3}a). $I$ increased in the lateral occipital cortex and in the inferior frontal gyrus (fig. \ref{Fig3}b).

Average effect sizes for relevant regions can be found in supplementary section 1. See supplementary section 2.2 for effects by onset site.

Some effects, for example in the supramarginal gyrus, were only visible in the independent variables, specifically in $K$. Some effects to cortical thickness, for example in the precentral gyrus, were not reflected in the independent morphological measures $K, I$, and $S$.

\subsection{Covariate effects}

We did not find any effects of age at surgery, duration of TLE, resection volume, side of resection, time between surgery and postoperative scan, or seizure outcome on the morphological changes following ATLR at a cluster significance of $0.05$.

\section{Discussion}

\subsection{Morphological findings}

We found significant changes in cortical morphology after ATLR in all measures we studied. The regions affected ipsilateral to the resection were the orbitofrontal and inferior frontal gyri, where we found changes to total surface area, exposed surface area, tension term, isometric term, and shape term, the lateral occipital gyrus, and lingual cortex, which saw changes in cortical thickness, exposed surface area, tension term, and isometric term. We also found thickness changes in parts of the ipsilateral pre- and postcentral gyri, as well as the supramarginal gyrus. Contralaterally, we found morphological changes in the occipital cortex in thickness, total surface area, tension term, and isometric term, and in the frontal cortex in total area, tension term, and isometric term. These findings indicate widespread structural cerebral changes after ATLR.

\subsection{Comparison to previous work}

Previous work has predominantly focused on changes to cortical thickness following ATLR. Our findings of changes in cortical thickness largely agree with those from previous studies; this includes decreases in thickness in the ipsilateral lateral occipital gyrus \citep{Elias2021} and in the ipsilateral lingual cortex \citep{Li2022}, and increased thickness in the ipsilateral precentral gyrus \citep{Zhao2021}. We additionally found cortical thickness increased in the contralateral lateral occipital gyrus. Changes to thickness can be offset by the surface area measures, explaining why studies might fail to replicate findings or even be in disagreement, showing how cortical thickness is not sufficient as a sole measure of cortical morphology. It may also explain why some changes in thickness were not reflected in the independent measures in our study.

\subsection{Interpretations}

Strong effects were seen in the orbitofrontal and inferior frontal gyri in almost all morphometric measures. Specifically, in the traditional morphological measures we saw increased total and exposed surface area, and in the independent measures an increase in the isometric term I, an increase in morphological complexity S, and a decrease in the tension term K. After surgery, the inferior and orbitofrontal cortex above the resection site often drops slightly into the resection cavity. Our observed effects are most likely explained by this “sagging” of the frontal lobe into the resection cavity; although, we cannot exclude that these effects are due to additional functional or structural mechanisms. We checked that these strong effects were not caused by distortion to the FreeSurfer region surface labels by visually inspecting the Desikan-Killiany atlas labelling in the post-surgery images. The thickness remaining constant in these gyri whilst the surface areas increase indicates that the cortical tissue is arranged differently in space, which is confirmed by changes to K and S. The decrease in $K$ supports the hypothesis that $K$ captures pressure on the cortex, which is reduced in the orbitofrontal and inferior frontal gyri because of the cavity.

In our study, the main regions with the greatest morphological changes after ATLR were structurally connected to the resected tissue. Previous studies on white matter tract alterations after temporal lobe surgery found reductions in quantitative anisotropy (QA) \citep{DaSilva2020} and fractional anisotropy (FA) \citep{Faber2013,Pustina2014,Concha2007,Winston2013,McDonald2010}  in the ipsilateral uncinate fasciculus, which connects the resected tissue to the orbitofrontal cortex, the region in which our analysis found the greatest morphological alterations. We also saw structural changes in the ipsilateral lateral occipital gyrus. This region is connected to the resected temporal lobe by the inferior longitudinal fasciculus, which has been shown to have reduced FA following surgery \citep{Faber2013,Pustina2014,Concha2007,Winston2013,McDonald2010}. There is evidence of reduced QA in the ipsilateral inferior fronto-occipital fasciculus \citep{DaSilva2020} and reduced FA in the ipsilateral inferior fronto-occipital fasciculus and the optic radiation \citep{Faber2013,Pustina2014,Concha2007,Winston2013,McDonald2010}, both passing through the temporal lobe to the occipital lobe. One study found increased FA in the contralateral uncinate fasciculus and superior longitudinal fasciculus \citep{Pustina2014}, evidence of white matter changes which could be underlying the morphological changes we found contralateral to the resection in the inferior frontal gyrus, frontal pole, and lateral occipital cortex. Previous studies also found alterations to the cingulum \citep{Faber2013,Winston2013,Concha2007,McDonald2010}. Although we did not find significant changes to the cingulate cortex in the traditional morphometrics, we were unable to investigate this region with independent morphometrics due to limitations of the method for their computation. Overall, our study suggests that the regions whose white matter connections are affected by surgery undergo structural changes. This could be due to Wallerian degeneration of the white matter tracts, a process in which an axon is cut or injured, causing parts of the axon distant from the damage to degenerate. This could in turn lead to atrophy in the connected grey matter, which is reflected in morphological changes. Future work on longitudinal structural, diffusion, and functional MRI will be necessary to confirm if this is in fact the underlying process.

None of the covariates we tested had significant effects on the morphological changes we found. In particular, there was no effect of the time since the first seizure occurred, indicating that the brains of those who experienced seizures for longer do not restructure differently than those with a short duration. Similarly, the age of the subject made no significant difference, although as subjects in our study were in an age range of 19 to 60 years, there might be age-dependent effects for individuals outside this range. Further, we did not see significant correlation with seizure outcome, despite evidence of larger white matter tract alterations in those who were seizure-free after surgery \citep{DaSilva2020}. This suggests that effects sizes of white matter and grey matter restructuring may not be correlated. Future work may explore additional outcome measures and covariates, such as language, visual, memory and cognitive function. Given that structural morphology relates to cognitive functioning in epilepsy  \citep{Garcia-Ramos2022}, investigating structural changes after surgery using independent components could help understand the mechanisms of, and treatment for, cognitive impairment.

We postulate that the structural changes we found are truly due to the surgery, rather than the progression of the disorder itself, for three reasons: First, although previous literature suggests that the progression of TLE affects cortical morphology, the changes described are not in regions such as the bilateral lateral occipital cortex or frontal poles \citep{Galovic2020,Whelan2018}, where we found changes between the pre- and postoperative scans. Additionally, these changes occur at longer time scales, with effects of up to 0.02mm reduction in cortical thickness per year \citep{Galovic2020,Whelan2018}. Thus, we would not expect strong effects over a relatively short period of two years. Second, we found no correlation between the duration of TLE and the morphological changes after ATLR. Third, the regions that we found to be affected by morphological changes are also connected to the resected temporal lobe by white matter tracts, suggesting a spatially specific non-random effect.

We cannot decisively determine if morphological change is a sign of atrophy or reorganisation, or potentially even increased function to compensate for the loss of tissue. A recent study on brain age changes after surgery found that an increased brain age in those with mTLE is reversed following surgery \citep{DeBezenac2021}. Similarly, progressive cortical thinning in unilateral TLE ceases after successful surgery to the rate of healthy ageing \citep{Galovic2020,Zhao2021}. These findings suggest that structural changes following surgery are linked to a restorative effect on brain health, which future work using independent morphometrics with longitudinal data may be able to verify. Of course, the processes of restructuring do not require spatial uniformity across the cortex, and there are likely region-specific effects with different causes. Further studies linking structural changes to postsurgical deficits such as visual field impairment will be necessary to deduce the nature of the processes causing the changes to morphology after ATLR. Furthermore, a focus on differences in morphological changes between individuals with persistent seizures after surgery and those rendered seizure free would also be valuable.

\subsection{Novel morphological measures}

We identified changes to cortical regions, such as the ipsilateral postcentral gyrus or the contralateral frontal pole, that the traditional measures did not find. This is in line with previous work \citep{Wang2021}, which showed that using the variables $K$, $I$, and $S$ can reveal morphological information otherwise concealed in the covariance between the measures thickness, exposed surface area, and total surface area. We found that the tension term $K$ detected even subtle morphological changes, likely due to its near-invariant value in controls, highlighting the benefits of independent measures for quantifying cortical shape.

\section{Conclusions}
We found widespread morphological changes following anterior temporal lobectomy, mainly in regions near the resections, but also remotely in regions that are structurally connected to the anterior temporal lobe, even contralaterally. This could be evidence of a reorganisation of the cortex after surgery, or atrophy caused by Wallerian degeneration of connected white matter structures. We found that the morphological effects were more pronounced and, in some cases, only detectable in a set of new, independent morphological measures of cortical morphology, rather than in traditional morphometric measures.

\section{Data and code availability}
Code for the computation of local morphological variables can be found on github: \url{https://github.com/KarolineLeiberg/folding_pointwise}.

\section{Acknowledgements}
We thank members of the Computational Neurology, Neuroscience \& Psychiatry Lab (www.cnnp-lab.com) for discussions on the analysis and manuscript. K.L. was supported by the EPSRC Centre for Doctoral Training in Cloud Computing for Big Data (EP/L015358/1). G.P.W. was supported by the MRC (G0802012, MR/M00841X/1). J.S.D., J.d.T., and S.B.V. are funded by UCL/UCLH and supported by the National Institute for Health and Care Research University College London Hospitals Biomedical Research Centre. P.N.T. and Y.W. are both supported by UKRI Future Leaders Fellowships (MR/T04294X/1, MR/V026569/1). B.M. is supported by Fundação Serrapilheira Institute (grant Serra-1709-16981) and CNPq (PQ 2017 312837/2017-8).

\section{Declaration of interest}
The authors declare no competing interests.

\bibliography{ResectionPaper}

\end{document}

\maketitle

\section{Average effect sizes of cortical regions} \label{effectSizes}

\begin{table}[h!]
\centering
\begin{tabular}{p{6cm} p{1cm} p{1cm} p{1cm} p{1cm} p{1cm} p{1cm}} 
 \hline \\ [0.05ex]
 Region & $T$ & $A_t$ & $A_e$ & $K$ & $I$ & $S$ \\ [2ex] 
 \hline \\ [0.05ex] 
 Ipsilateral inferior frontal gyrus & 0 & \textbf{0.20} & \textbf{0.35} & \textbf{-0.91} & \textbf{0.29} & \textbf{0.25} \\ [1.5ex]
 Ipsilateral orbitofrontal gyrus & -0.07 & \textbf{0.29} & \textbf{0.58} & \textbf{-2.38} & \textbf{0.43} & \textbf{0.39} \\ [1.5ex]
 Ipsilateral precentral gyrus & \textbf{0.06} & -0.04 & 0.02 & \textbf{-0.08} & 0 & -0.04 \\ [1.5ex]
 Ipsilateral postcentral gyrus & 0.06 & 0.07 & 0.06 & \textbf{0.03} & 0.07 & 0.06 \\ [1.5ex]
 Ipsilateral supramarginal gyrus & 0 & -0.08 & -0.09 & \textbf{0.10} & -0.08 & -0.08 \\ [1.5ex]
 Ipsilateral lateral occipital cortex & \textbf{-0.07} & -0.02 & -0.07 & \textbf{0.09} & \textbf{-0.05} & -0.05 \\ [1.5ex]
 Ipsilateral lingual cortex & \textbf{-0.09} & -0.08 & \textbf{-0.10} & 0.02 & -0.08 & -0.05 \\ [1.5ex]
 Contralateral lateral occipital cortex & \textbf{0.06} & .08 & 0.08 & \textbf{0.07} & \textbf{0.09} & 0.05 \\ [1.5ex]
 Contralateral inferior frontal gyrus & 0.05 & \textbf{0.07} & 0.06 & 0.08 & \textbf{0.07} & 0 \\ [1.5ex]
 Contralateral frontal pole & 0 & \textbf{0.06} & 0 & \textbf{0.08} & 0.06 & 0 \\ [2ex]
 \hline
\end{tabular}
\caption{Average effect sizes $f^2$. Decreases are shown as negative effects. Bold font indicates the cluster effect was statistically significant at 5\%.}
\label{tableEffects}
\end{table}

\newpage

\section{Results by seizure onset group}

\subsection{Traditional morphological measures} \label{rawLR}

\begin{figure}[htp]
\centering
\includegraphics[width=16.5cm]{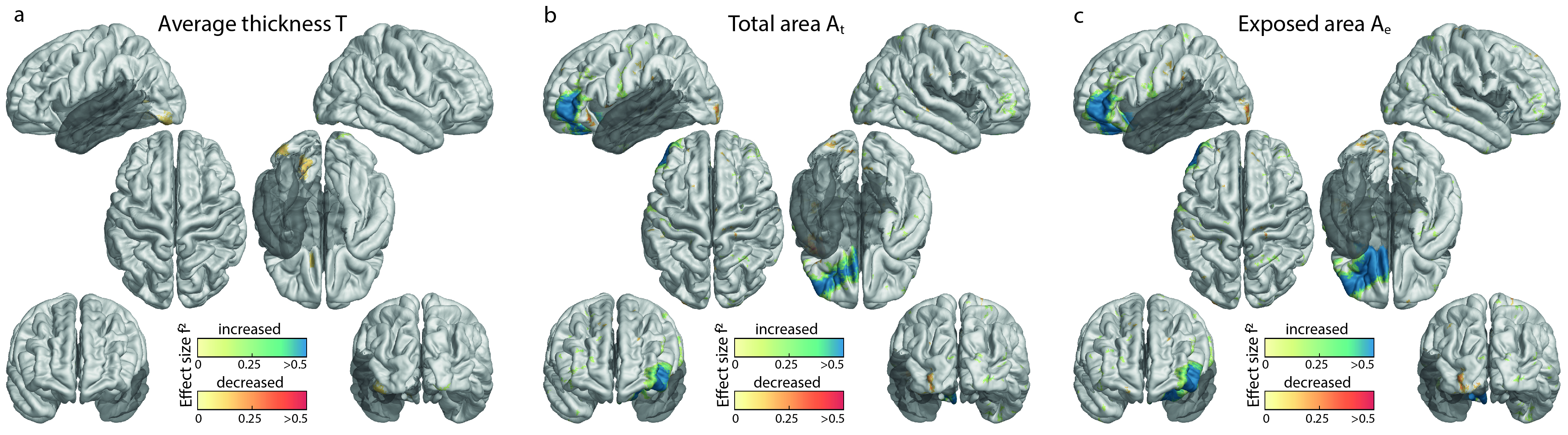}
\caption{Changes following ATLR in individuals with left onset in the morphological variables average cortical thickness $T$ (a), total surface Area $A_t$ (b), and exposed surface area $A_e$ (c). Effect clusters with statistical significance of $p \leq 0.05$ are shown and effect sizes are given in Cohen’s $f^2$ metric. Increases are in blue, decreases are in red. Regions excluded from the analysis are in dark grey.} \label{rawLH}
\end{figure}

\begin{figure}[htp]
\centering
\includegraphics[width=16.5cm]{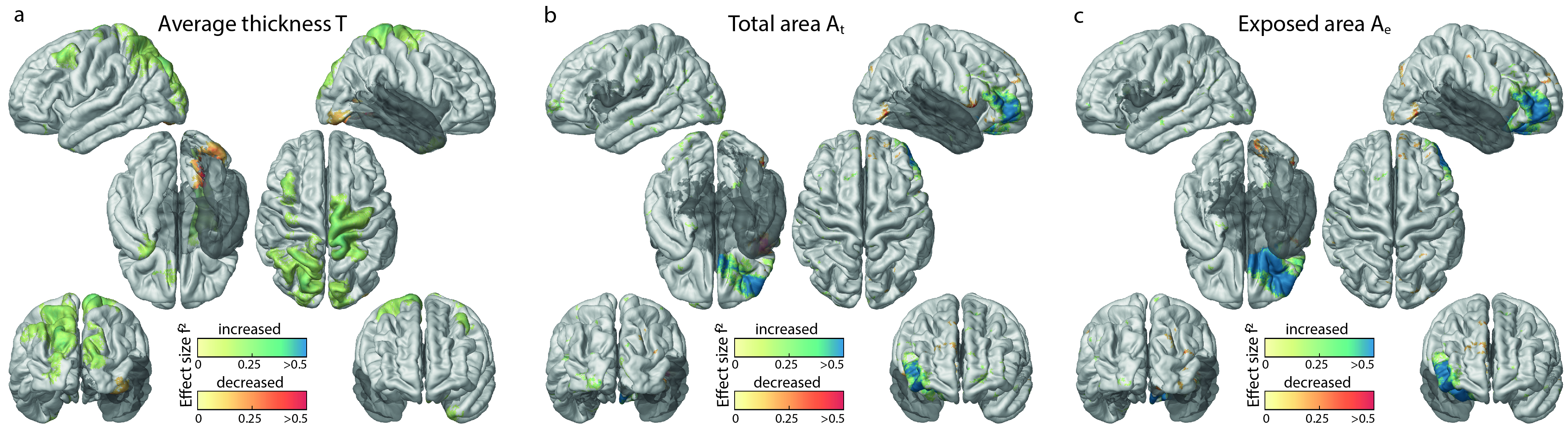}
\caption{Changes following ATLR in individuals with right onset in the morphological variables average cortical thickness $T$ (a), total surface Area $A_t$ (b), and exposed surface area $A_e$ (c). Effect clusters with statistical significance of $p \leq 0.05$ are shown and effect sizes are given in Cohen’s $f^2$ metric. Increases are in blue, decreases are in red. Regions excluded from the analysis are in dark grey.} \label{rawRH}
\end{figure}

Although there are more significant changes to average cortical thickness in individuals with right resections (fig. \ref{rawRH}a) than left resection (fig. \ref{rawLH}a), we did not find significant effects in a direct comparison of left and right onset. This might be due to the sample size of one group being not large enough or its variance being too high to reveal any significant clusters when testing for an effects of onset side.

\newpage

\subsection{Independent morphological measures} \label{kisLR}

\begin{figure}[htp]
\centering
\includegraphics[width=16.5cm]{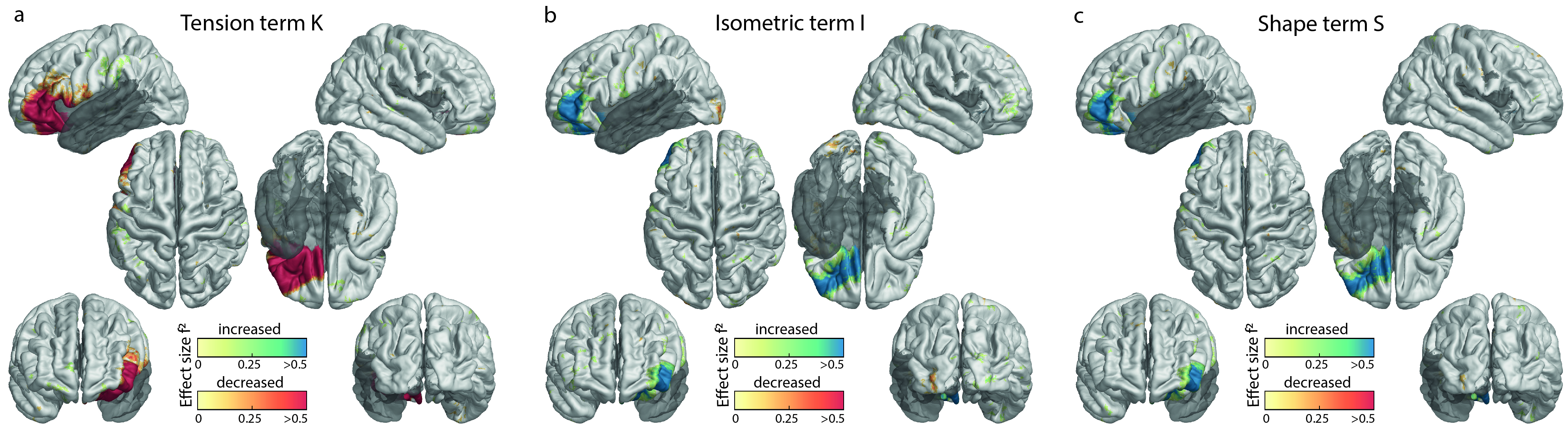}
\caption{Changes following ATLR in individuals with left onset in the independent morphological measures $K$ (a), $I$ (b), and $S$ (c). Effect clusters with statistical significance of p<0.05 are shown and effect sizes are given in Cohen’s $f^2$ metric. Increases are in blue, decreases are in red. Regions excluded from the analysis are in dark grey.} \label{kisLH}
\end{figure}

\begin{figure}[htp]
\centering
\includegraphics[width=16.5cm]{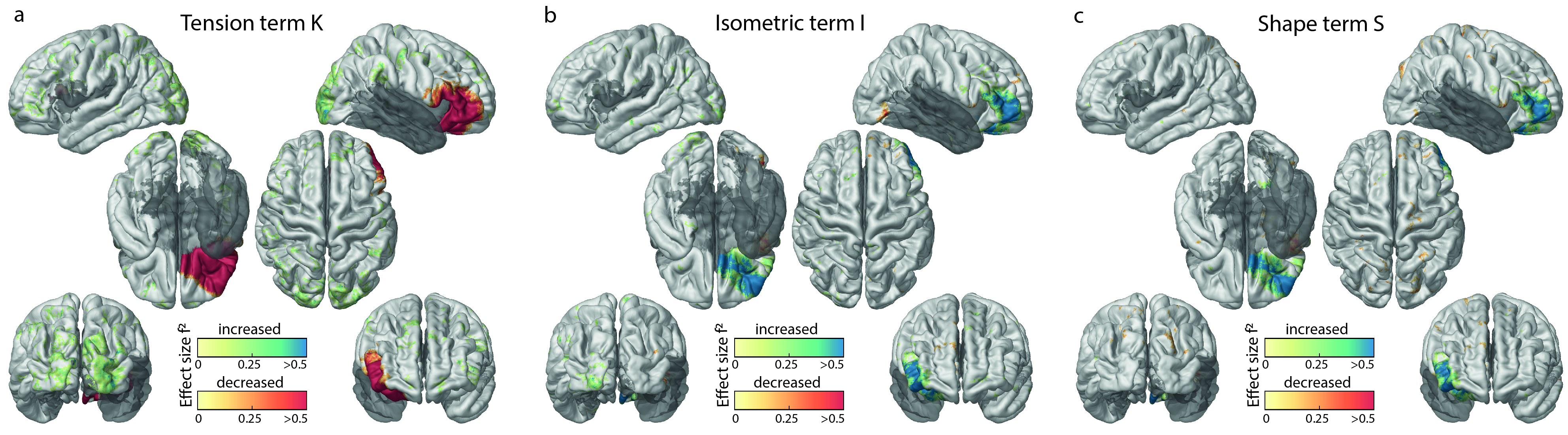}
\caption{Changes following ATLR in individuals with right onset in the independent morphological measures $K$ (a), $I$ (b), and $S$ (c). Effect clusters with statistical significance of p<0.05 are shown and effect sizes are given in Cohen’s $f^2$ metric. Increases are in blue, decreases are in red. Regions excluded from the analysis are in dark grey.} \label{kisRH}
\end{figure}